\documentclass[a4paper,10pt,twoside]{cpc-hepnp}

\usepackage{multicol}
\usepackage{graphicx}
\usepackage{booktabs}
\usepackage{amssymb,bm,mathrsfs,bbm,amscd}
\usepackage[tbtags]{amsmath}
\usepackage{lastpage}
\usepackage{CJK}
\usepackage[english]{babel}
\usepackage{multirow}
\usepackage{array}
\usepackage{url}

\begin{document}
\begin{CJK*}{GBK}{song}

\fancyhead[c]{\small Chinese Physics C~~~Vol. XX, No. X (201X) XXXXXX} \fancyfoot[C]{\small XXXXXX-\thepage}

\footnotetext[0]{Received XX XXXX 201X}

\title{Development of Yangbajing Air shower Core detector array for a new EAS hybrid Experiment\thanks{Supported by the Grants from the National Natural Science Foundation of China (Y11122005B, Y31136005C, Y0293900TF and 11165013) and the Chinese Academy of Sciences (H9291450S3 and Y4293211S5) and the Knowledge Innovation Fund of Institute of High Energy Physics(IHEP), China (H95451D0U2 and H8515530U1)}}

\author{%
\quad LIU Jin-Sheng$^{1;1}$\email{liujs@ihep.ac.cn}%
\quad HUANG Jing$^{1}$%
\quad CHEN Ding$^{2}$
\quad ZHZNG Ying$^{1}$\\
\quad ZHAI Liu-Ming$^{1}$
\quad CHEN Xu$^{1}$
\quad HU Xiao-Bin$^{1,3}$
\quad LIN Yu-Hui$^{1}$\\
\quad ZHANG Xue-Yao)$^{3}$
\quad FENG Cun-Feng$^{3}$
\quad JIA Huan-Yu$^{4}$\\
\quad ZHOU Xun-Xiu$^{4}$
\quad DANZENGLUOBU$^{5}$
\quad CHEN Tian-Lu$^{5}$\\
\quad LI Hai-Jin$^{5}$
\quad LIU Mao-Yuan$^{5}$
\quad YUAN Ai-Fang$^{5}$
}
\maketitle

\address{$^1$ Key Laboratory of Particle Astrophysics, Institute of High Energy Physics, Chinese Academy of Sciences, Beijing 100049, China}
\address{$^2$ National Astronomical Observatories, Chinese Academy of Sciences, Beijing 100012, China}
\address{$^3$ Department of Physics, Shandong University, Jinan 250100, China}
\address{$^4$ Institute of Modern Physics, Southwest Jiaotong University, Chengdu 610031, China}
\address{$^5$ Physics Department of Science School, Tibet University, Lhasa 850000, China}

\begin{abstract}
  Aiming at the observation of cosmic-ray chemical composition at the ``knee" energy region, we have been developing
  a new type air-shower core detector (YAC, Yangbajing Air shower Core detector array) to be  set up at Yangbajing (90.522$^\circ$ E, 30.102$^\circ$ N, 4300 m above sea level, atmospheric depth: 606 g/m$^2$) in Tibet, China. YAC works together with the Tibet air-shower array (Tibet-III) and an underground water cherenkov muon detector array (MD) as a hybrid experiment. Each YAC detector unit consists of lead plates of 3.5 cm thick and a scintillation counter which detects the burst size induced by high energy particles in the air-shower cores. The burst size can be measured from 1 MIP (Minimum Ionization Particle) to $10^{6}$ MIPs. The first phase of this experiment, named ``YAC-I", consists of 16 YAC detectors each having the size 40 cm $\times$ 50 cm and distributing in a grid with an effective area of 10 m$^{2}$. YAC-I is used to check hadronic interaction models. The second phase of the experiment, called ``YAC-II", consists of 124 YAC detectors with coverage about 500 m$^2$. The inner 100 detectors of 80 cm $\times $ 50 cm each are deployed in a 10 $\times$ 10 matrix from with a 1.9 m separation and the outer 24 detectors of 100 cm $\times$ 50 cm each are distributed around them to reject non-core events whose shower cores are far from the YAC-II array. YAC-II is used to study the primary cosmic-ray composition, in particular, to obtain the energy spectra of proton, helium and iron nuclei between 5$\times$$10^{13}$ eV and $10^{16}$ eV covering the ``knee" and also being connected with direct observations
  at energies around 100 TeV. We present the design and performance of YAC-II in this paper.
\end{abstract}



\footnotetext[0]{\hspace*{-3mm}\raisebox{0.3ex}{$\scriptstyle\copyright$}2013
Chinese Physical Society and the Institute of High Energy Physics
of the Chinese Academy of Sciences and the Institute
of Modern Physics of the Chinese Academy of Sciences and IOP Publishing Ltd}%

\begin{multicols}{2}

\section{Introduction}
\label{intro}
The all-particle energy spectrum of primary cosmic rays is well expressed by a power law dN/dE $\propto$ E$^{-\gamma}$ over many orders of magnitude, with $\gamma$ changing sharply from 2.7 to 3.1 at about 4 PeV~\cite{Amenomori-2008}. The break of the all-particle energy spectrum is called the ``knee"~\cite{Horandel-2003,JHuang-2003}, and the corresponding energy range ($10^{15}$ eV - $10^{16}$ eV) is called ``knee region". Although the existence of the knee is well confirmed by many experiments, there still exist debates on its origin. In order to clarify the sharpness of the knee, precise measurements as possible of the primary spectra of individual components including the knee will be essentially important. The best ways to study the chemical compositions are direct measurements of primary cosmic rays on board balloons or satellites, but its
energy range with sufficient statistics is limited to 10$^{14}$ eV because of limited exposure time and small detective area. So the task of studying chemical components of knee region still relies on the ground-based indirect measurements.

The study of the mass composition of primary cosmic rays around the knee has been made in the last decade by the Tibet-EC (emulsion chamber) experiment using high threshold ($>$ TeV) air-shower core detector sensitive to the primary particle mass~\cite{Amenomori-2006}. It was aimed to separate air-shower events induced by primary light component such as protons and helium nuclei. The observation shows low intensities of proton and helium spectra which amount to less than 30\% of all-particle spectrum, suggesting that the knee is dominated by nuclei heavier than helium. Demerits of this experiment are: 1)the statistics are limited due to the selection of high-energy core events; 2)the reconstruction of the primary energy spectrum is based on the AS simulations in which the hadronic interaction model is not fully established yet, although the model dependence of AS core events at high observation level is at most few tens percent as already reported~\cite{Amenomori-2006}. To overcome these problems, we have recently made the upgrade of the Tibet-EC experiment and started a new low threshold core detector named YAC (Yangbajing Air shower Core detector), which has a capability of observing the core events with high statistics as well as testing the interaction models.

 In order to further expand the energy region of the Tibet-EC experiment, YAC detector has been developed to lower the threshold energy as possible with wider dynamic range. The detection threshold energy of the YAC is able to set at several 10 times lower than EC (about 300 GeV corresponding to the primary energy of several times 10 TeV) by adopting the scintillator instead of the X-ray film for the detection of cascade showers induced in the lead plate by high energy AS core particles, and the wide dynamic range of 1-$10^{6}$ MIPs (Minimum Ionization Particles) for the burst size detection is realized by installing 2 PMTs (high-gain and low-gain PMTs). Using the wave length shifting fiber to collect the scintillating light improves the geometrical uniformity. This new experimental condition improves the statistics of the high energy core events by a factor of 100, when compared with the Tibet-EC experiment . The new hybrid experiment aims to observe the energy spectra of proton, helium and iron whose energy range will overlap with direct observations at lower energies such as CREAM~\cite{CREAM3-phe,CREAM3-other} and ATIC~\cite{ATIC2-2005}, and Tibet-EC experiment at higher energies. Furthermore, we may add the underground muon detector (MD) to this experiment to increase the mass separation power of primary particles. Hence, the new Tibet hybrid experiment consists of YAC array, Tibet-III array~\cite{Amenomori-2008} and the underground MD array (Fig.~\ref{figure-01}).

The design of the YAC detector is described in section 2. YAC experiment is scheduled in two steps, named ``YAC-I" and ``YAC-II". YAC-I is a small array consisting of 16 prototype detectors covering area of about 10 m$^2$ located near the center of the Tibet-III array. This array will operate for a few months to observe the AS core events of primary energies around $10^{14}$ eV where the mass composition of primary cosmic rays is fairly known by direct observations~\cite{CREAM3-phe,CREAM3-other,ATIC2-2005}. Therefore, the role of YAC-I is to test the interaction models
currently used by Monte Carlo (MC) simulations such as QGSJET01~\cite{qgs01}, SIBYLL2.1~\cite{siby}, DPMJET~\cite{dpmjet} and lately EPOS-LHC~\cite{eposlhc}, QGSJETII-04~\cite{qgs-2-4}. YAC-II is an array of 124 detectors covering area of $\sim$500 m$^2$ to obtain proton, helium and iron spectra with high statistics in energy range between 5$\times$$10^{13}$ eV and $10^{16}$ eV, which will be smoothly connected to those by directs observations in lower energy region. In this paper, the design of YAC-II and its performance are described.

%\end{multicols}
%\ruleup
%\begin{center}
%\includegraphics[width=12cm]{cpcf2.eps}
%\figcaption{\label{fig2} Figure 2.}
%\end{center}
%\ruledown
%\begin{multicols}{2}

\section{YAC-II Experiment}
\label{exp}
\subsection{The design of YAC-II array}

The YAC array is set up at Yangbajing (90.522$^\circ$ E, 30.102$^\circ$ N, 4300 m above sea level, atmospheric depth: 606 g/m$^2$) in Tibet, China. This array works together with the Tibet-III and underground MD array as a hybrid experiment, as shown in Fig.~\ref{figure-01}.

\begin{center}
\includegraphics[width=0.8\linewidth]{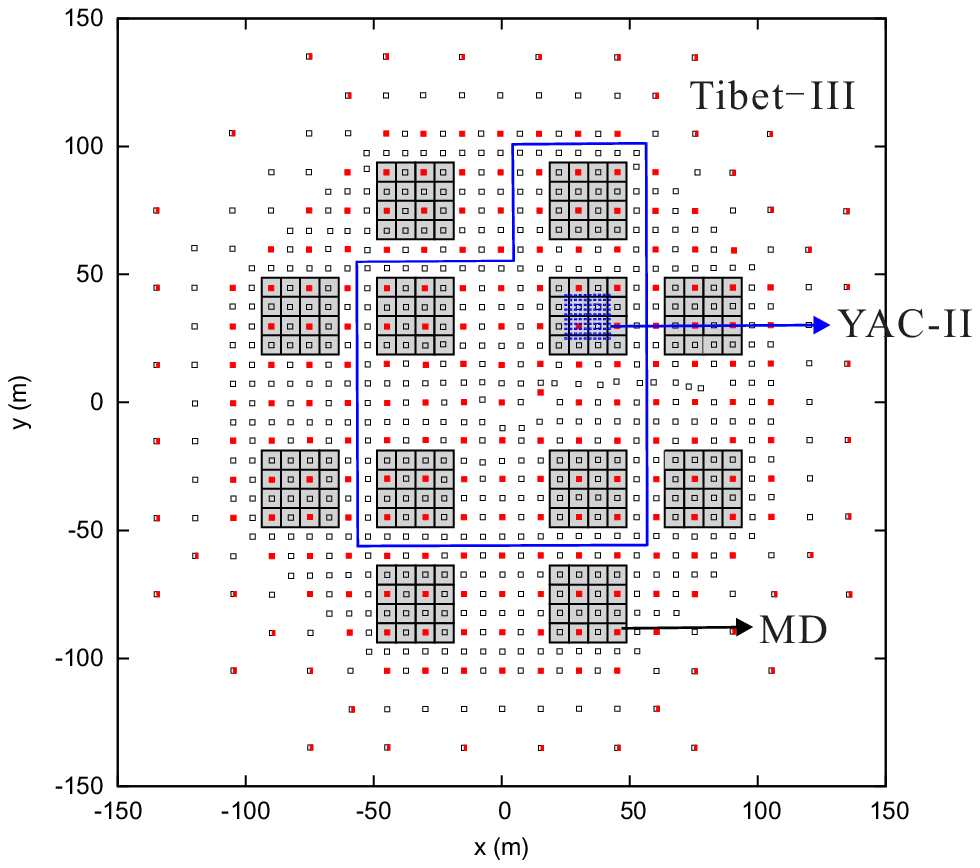}
\figcaption{Schematic view of (YAC-II+Tibet-III+MD) array. The YAC-II array contains 124 core detectors (blue open squares) and is located at near the center of Tibet-III ($\sim$50000 m$^2$). The Tibet-III array consists of 576 fast timing (FT) detectors (black open squares), 28 density (D) detectors (red half open squares) and 185 FT/D detectors (red solid squares) at periphery. The MD array consists of 12 underground water pools (gray solid squares), each of which has 16 units. 5 water pools ($\sim$4500 m$^2$) in blue wire frame have been built.}
\label{figure-01}
\end{center}

YAC-II consists of 124 core detectors covering the area of about 500 m$^2$, as shown in Fig.~\ref{figure-02}. The inner 100 detectors 80 cm $\times $ 50 cm each are deployed in a 10 $\times$ 10 matrix from with a 1.9 m separation and the outer 24 detectors 100 cm $\times$ 50 cm each are distributed around them to reject non-core events whose shower cores are far from the YAC-II array. The main purpose of YAC-II is to observe the energy spectra of light component as well as iron nuclei in the primary cosmic rays between 5$\times$$10^{13}$ eV and $10^{16}$ eV as mentioned in Section 1
\begin{center}
\includegraphics[width=0.8\linewidth]{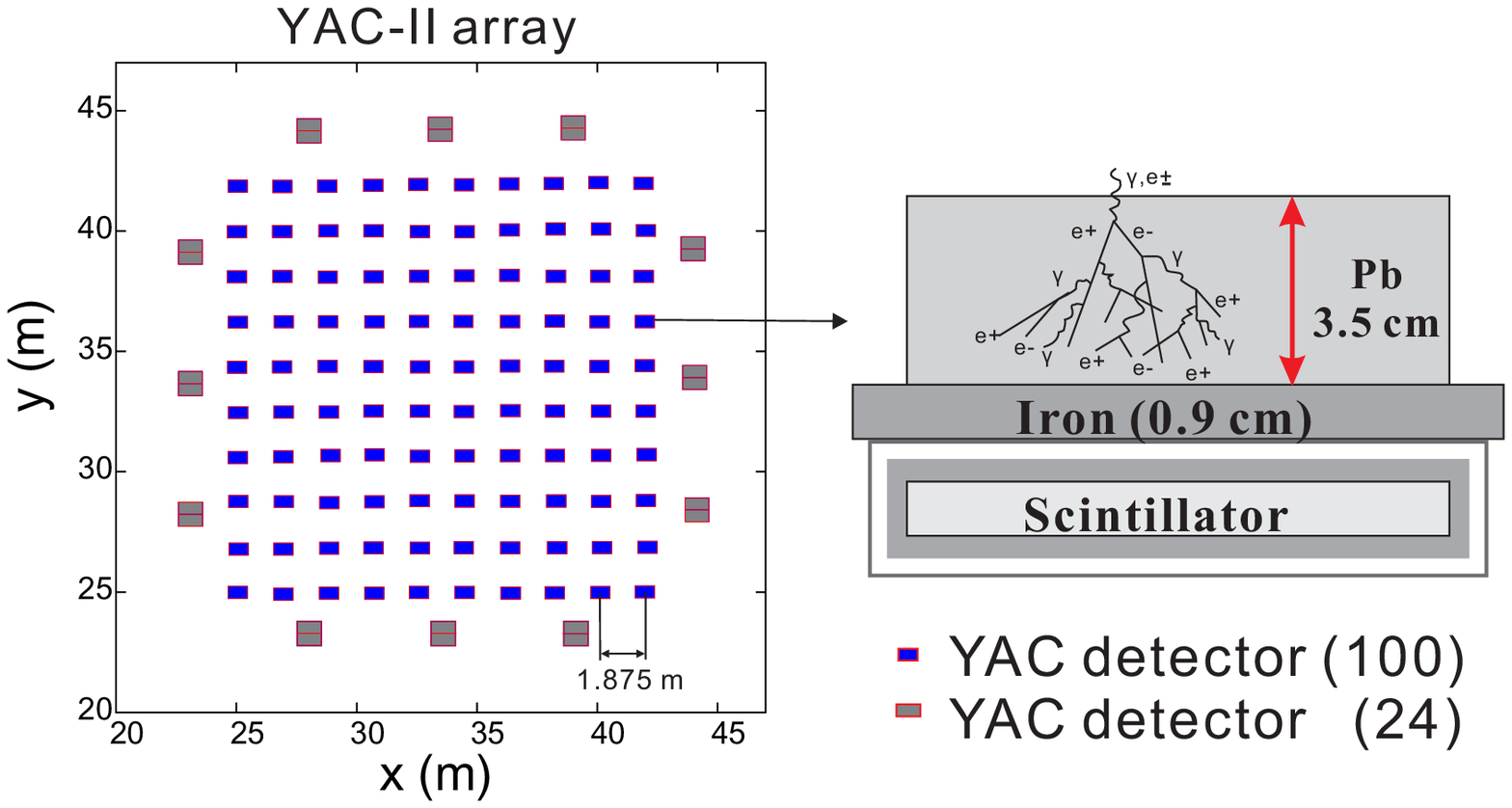}
\figcaption{Left: The magnified schematic view of the YAC-II array ($\sim$500 m$^2$). The detectors of this array are classified to 2 types: the inner 100 detectors of 0.4 m$^2$ each (blue solid squares) and the outer 24 detectors of 0.5 m$^2$ each (gray solid squares). Right: The schematic view of each detector comprising a lead plate of 3.5 cm thickness, a iron plate of 0.9 cm thickness and a plastic scintillator of 1 cm thickness.}
\label{figure-02}
\end{center}

\subsection{The design of YAC-II detector}

The design of YAC-II detector essentially followed that described in~\cite{Katayose-2005} but in this work some improvements were made as follows:

(1) Lead plates: A lead layer is used to select high energy particles in the AS core in the energy range from several GeV to several 10 TeV, as shown in Fig.~\ref{figure-02}. It is found from an optimization calculation that the lead layer of thickness 3.5 cm (6.3 radiation length) meets the measurements in both lower and higher energies. A 0.9 cm thick iron plate placed in the middle of scintillator and lead plates is used to support the weight of lead plates, as shown in Fig.~\ref{figure-02}. High energy electromagnetic particles near the AS axis develop into cascade showers in the
lead absorber and these shower particles enter to the scintillator counter. Here, the number of shower particles detected in the scintillator is defined as the burst size ($N_b$). The scintillation light produced by shower electrons in the scintillator below the lead layer is transmitted to PMTs via the wave length shifting fibers (WLSF BCF-92, SAINT-GOBAIN, round cross section with 1.5 mm diameter), as shown in Fig.~\ref{figure-03}. The determination of the burst size ($N_{b}$) is calibrated by using a charge count value of single-muon peak.

\begin{center}
\includegraphics[width=0.7\linewidth]{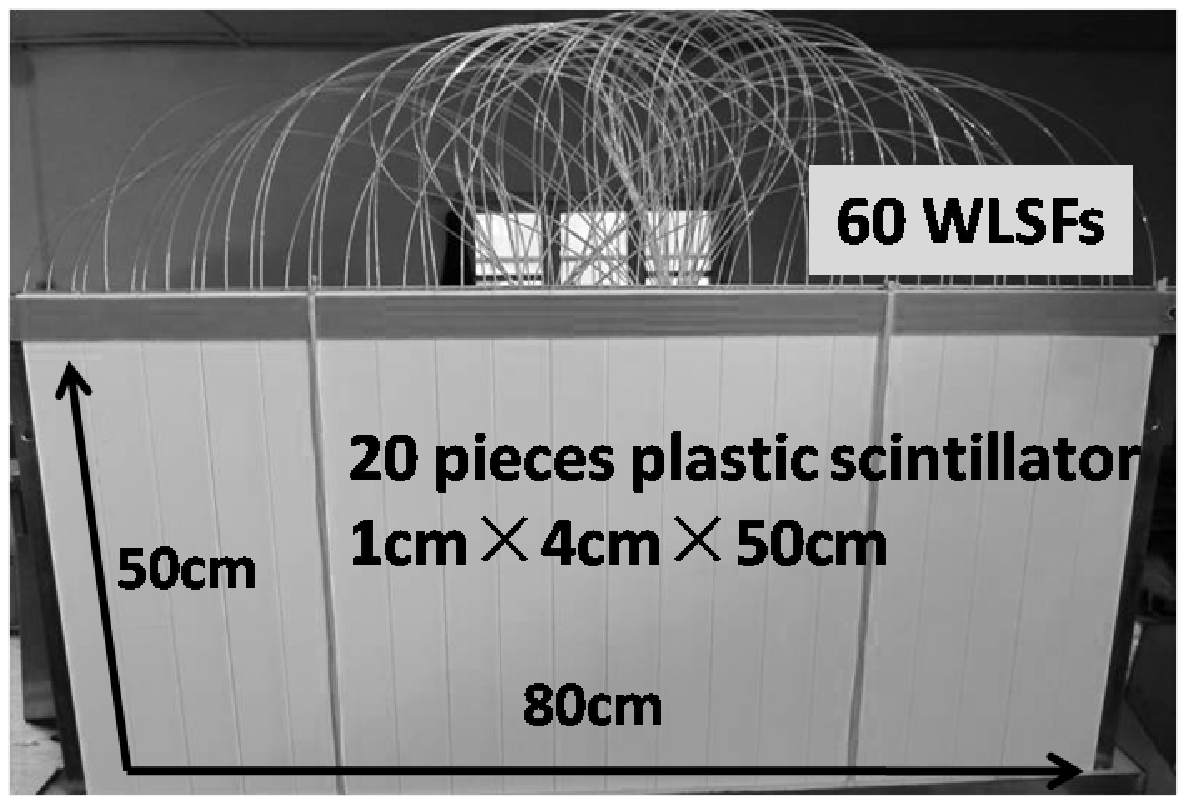}
\includegraphics[width=0.7\linewidth]{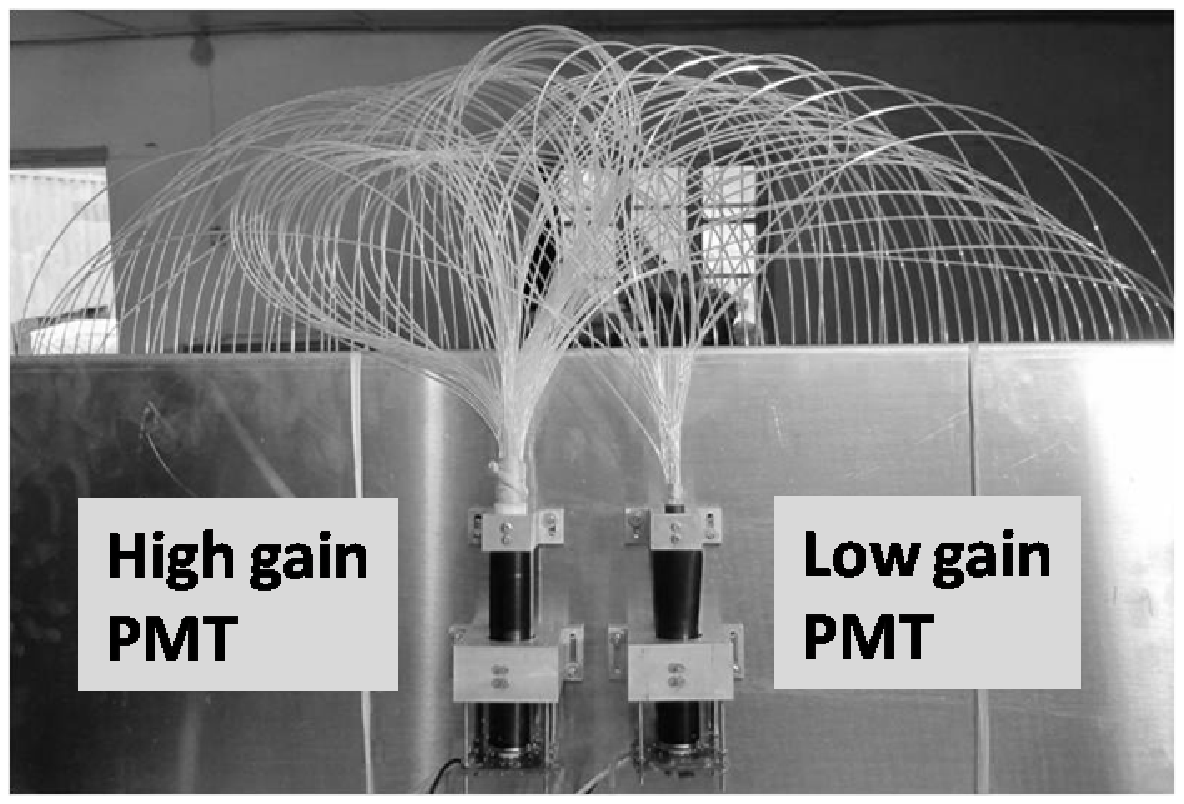}
\figcaption{The view of the scintillation detector. Top: the top view and Bottom: the back view. The scintillation detector unit consists of 20 pieces scintillator whose size is 50 cm $\times$ 4 cm $\times$ 1 cm and light-isolated from each other using reflecting material. All the fibers are adjusted to the same length of 110 cm. Both of 20 fibers attached to the low-gain PMT (R5325) and 40 fibers attached to high-gain PMT (R4125) are installed in the center of the scintillator with equal distance.}
\label{figure-03}
\end{center}

(2) Scintillation counter: In order to have a certain position resolution of high energy particles in the air shower cores, the size of an inner detector unit is taken to be 80 cm $\times$ 50 cm, as shown in Fig.~\ref{figure-03}. One unit scintillation counter is divided into 20 pieces of 4 cm width. For getting a better uniformity of light output when a high energy particle hits different position of a scintillation counter unit, 40 WLSFs are installed parallelly in the center of each scintillator and connected to high-gain PMT (R4125) while 20 WLSFs installed in the center of each scintillator are connected to low-gain PMT (R5325). Such design guarantees geometrical uniformity of the detector response within 6\%.

(3) PMT: In order to record the electromagnetic showers in the energy range from several GeV to  several 10 TeV, a wide dynamic range from 10 MIPs to $10^6$ MIPs of PMT is required. In addition, taking into account the importance of single-particle measurement in the system calibration, the dynamic range of PMT should be from 1 MIP to $10^6$ MIPs. This is realized by adopting a high-gain PMT (HG-PMT, HAMAMATSU R4125) and a low-gain PMT (LG-PMT, HAMAMATSU R5325) that are responsible for the range of $1- $$3\times10^3$ MIPs and $10^3-10^6$ MIPs, respectively.

(4) WLSFs: To save the wave length shifting fibers, a short length (110 cm, half of the fiber length used in the old detectors~\cite{Katayose-2005}) of fiber is used for which one end is connected with the PMT and another end is plated with aluminum for the reflection, as shown in Fig.~\ref{figure-03}. Since the reflection rate reaches 99\% and only the gain of PMT is recorded from each detector, there is no influence on the experimental data. Considering the large
temperature difference in Yangbajing within one year even one day, a heat-insulator layer is used inside each detector unit box.

YAC-II array has been operating together with the Tibet-III array and the underground MD array since March, 2014, as shown in Fig.~\ref{figure-01}. The Tibet-III array is used to measure the shower size and the arrival direction of each air shower. Any four-fold coincidence of the FT detectors is used as the trigger condition for air-shower events. The air-shower direction can be estimated with an error smaller than 0.2$^{0}$ above 100 TeV~\cite{Amenomori-2008}, and the primary energy resolution is estimated to be 12\% at energies around 10$^{15}$ eV by our simulation~\cite{YAC-huang}.

Trigger rate of YAC-II is about 3.5 Hz with a dead-time rate of $\sim$ 1\%. If one YAC-II detector makes a trigger signal,
all ADC data from all YAC-II units are recorded. Also the trigger signal is sent to DAQ system for Tibet-III array
and MD array. ADC modules of YAC-II are calibrated every 4 hours. ADC pedestal values  are measured every 10 minutes.
Each DAQ system has a GPS clock module independently. The matching between YAC, Tibet-III and MD data is made
by taking a coincidence of GPS clocks, trigger tag to Tibet-III and MD array. The coincidence interval of GPS is shorter than 1 ms. The average time difference is 8.1 $\mu$s $\pm$ 0.4 $\mu$s.

\section{Detector Calibration}
\label{performance}
\subsection{The probe calibration}
In this hybrid experiment, the burst size is defined as the PMT output (charge) divided by that of the single-particle peak, which is determined by a probe calibration using cosmic rays, typically  muons. For this purpose, a small scintillator 25 cm $\times$ 25 cm $\times$ 3.5 cm with a PMT (H1949) is put on the top of each detector during the maintenance period. This is called a probe detector, as shown in Fig.~\ref{figure-04}. Fig.~\ref{figure-05} shows the charge distribution of single particle in a detector unit. The peak is defined as one MIP ( 1 MIP = 2.12 $\pm$ 0.01 pC ). Therefore, the determination of the burst size $N_{b}$ is calibrated by using a single-particle peak.

\begin{center}
\includegraphics[width=0.65\linewidth]{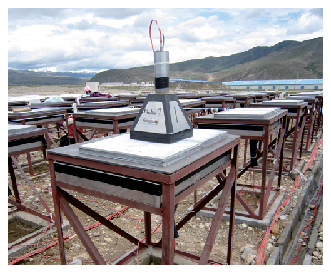}
\figcaption{The view of the probe calibration.}
\label{figure-04}
\end{center}
\begin{center}
\includegraphics[width=0.8\linewidth]{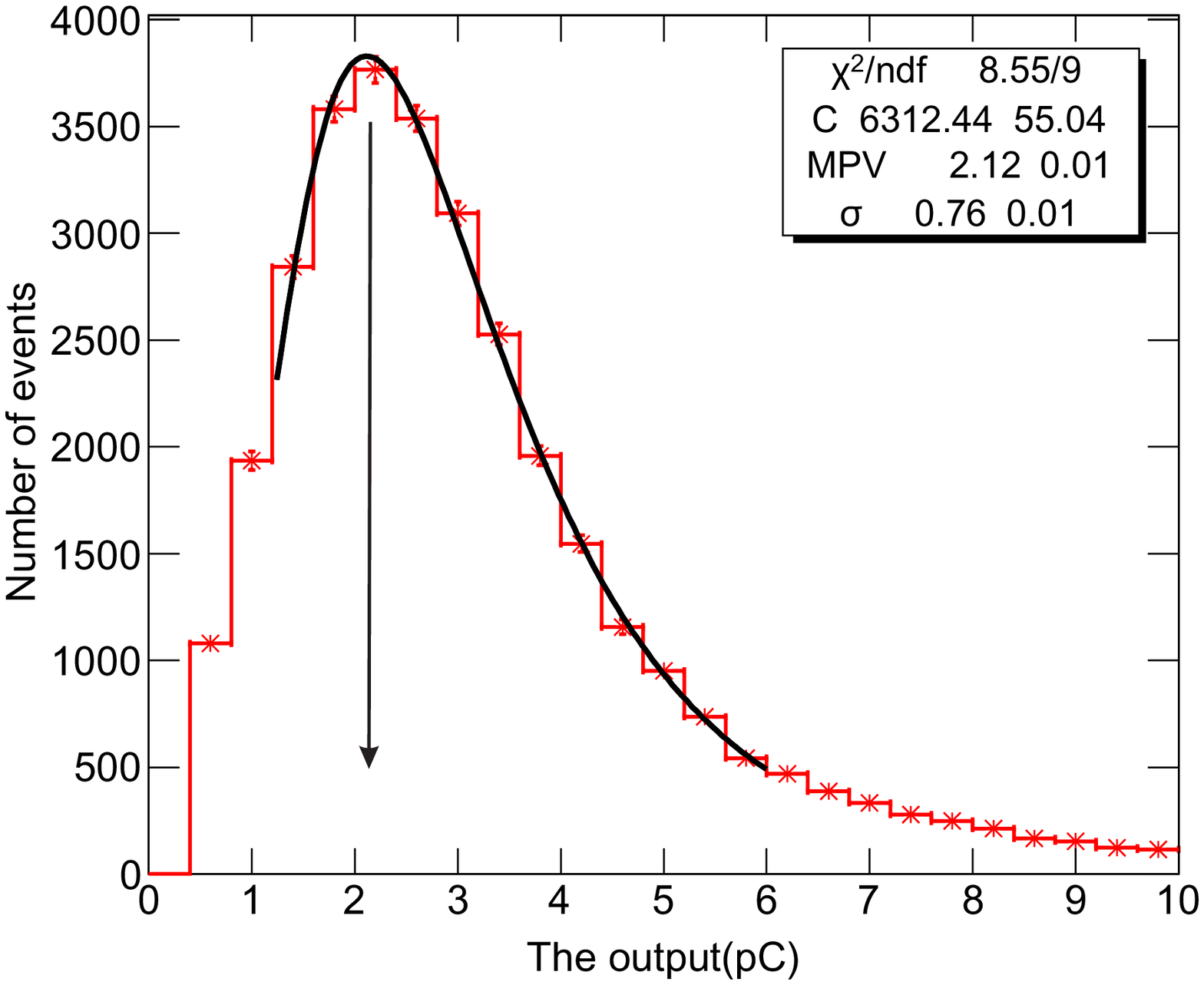}
\figcaption{The charge distribution of a single muon in the first detector unit. The peak called single-particle peak is defined as one MIP. The single-particle peak of this detector is 2.12 $\pm$ 0.01 pC.}
\label{figure-05}
\end{center}

\subsection{Linearity of PMTs}
 For each PMT (R4125 and R5325) used in YAC-II, the linearity with respect to high voltage is better than 1\%, as shown in Fig.~\ref{figure-06}. The high-gain PMT (R4125) is used to measure light yield from 1 MIP to a few thousand MIPs, while low-gain PMT (R5325) cover the dynamic range from a few hundred to $10^{6}$ MIPs. The dynamic range of PMTs is measured using LED light source and optical filters. In the test experiment we fixed the positions of LED, filter and PMT. By using different filters, we can get lights of the different intensity. The LED is driven by TTL pulse with the width of 25 nsec.
The details are described in ~\cite{Katayose-2005}.

\subsection{Uniformity of the detector}
When the charged particles pass through the scintillator in different position, the light transmitted to the PMT by the WLSFs will vary in a little range which is called ``position dependence" or ``uniformity" of the detector. The uniformity of YAC-II detector has been measured using cosmic-ray single muons selected by a triple coincidence as shown in Fig.~\ref{figure-07}. As shown in the figure, the uniformity in the scintillator length 50 cm is better than 6$\%$.

\begin{center}
\includegraphics[width=0.8\linewidth]{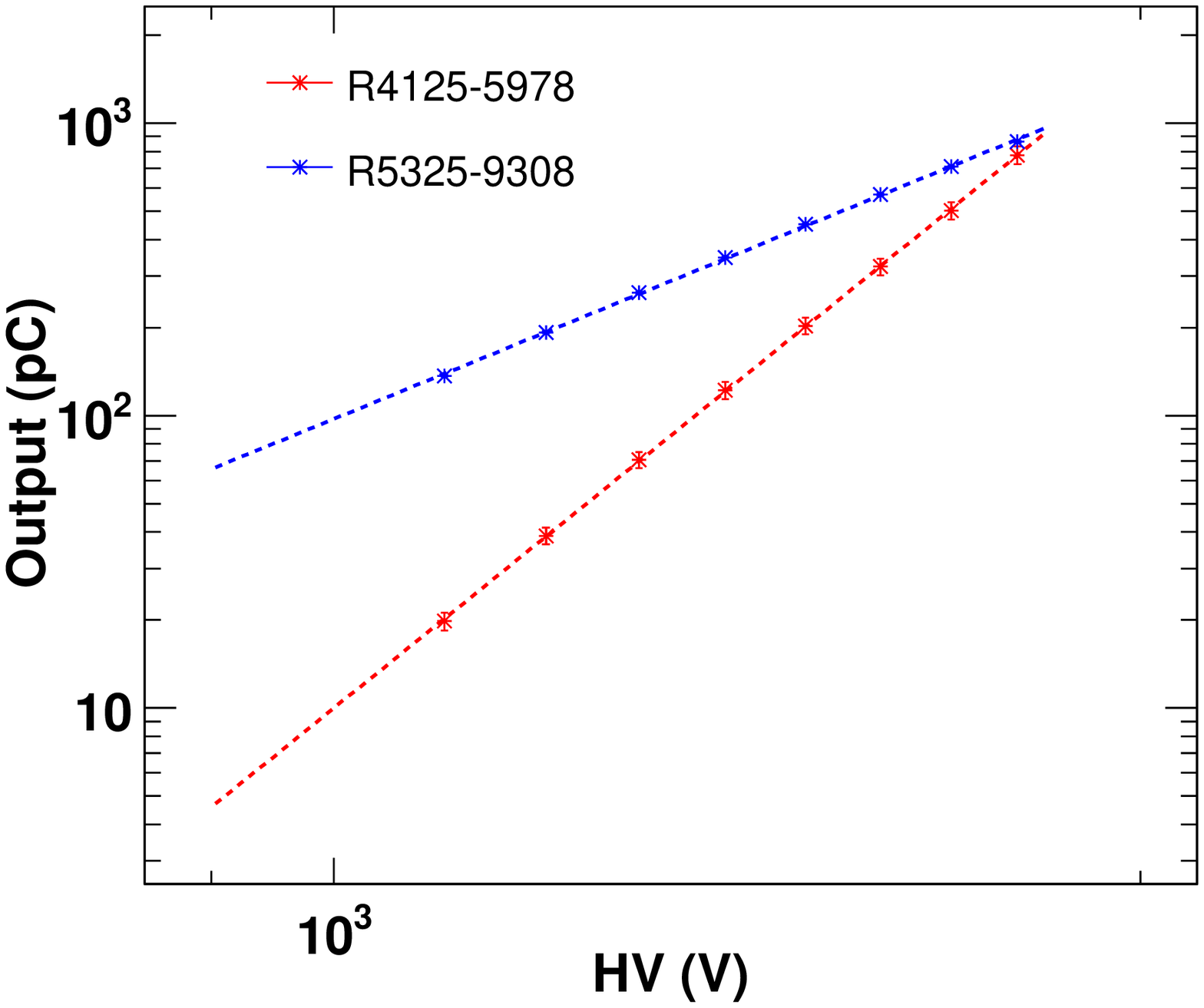}
\figcaption{The linearity between the output (pC) and the high voltage supplied to high-gain PMT (R4125, Red) and low-gain PMT (R5325, Blue). The points show the experimental data. The red dotted line is a fitting line.}
\label{figure-06}
\end{center}

\begin{center}
\includegraphics[width=0.8\linewidth]{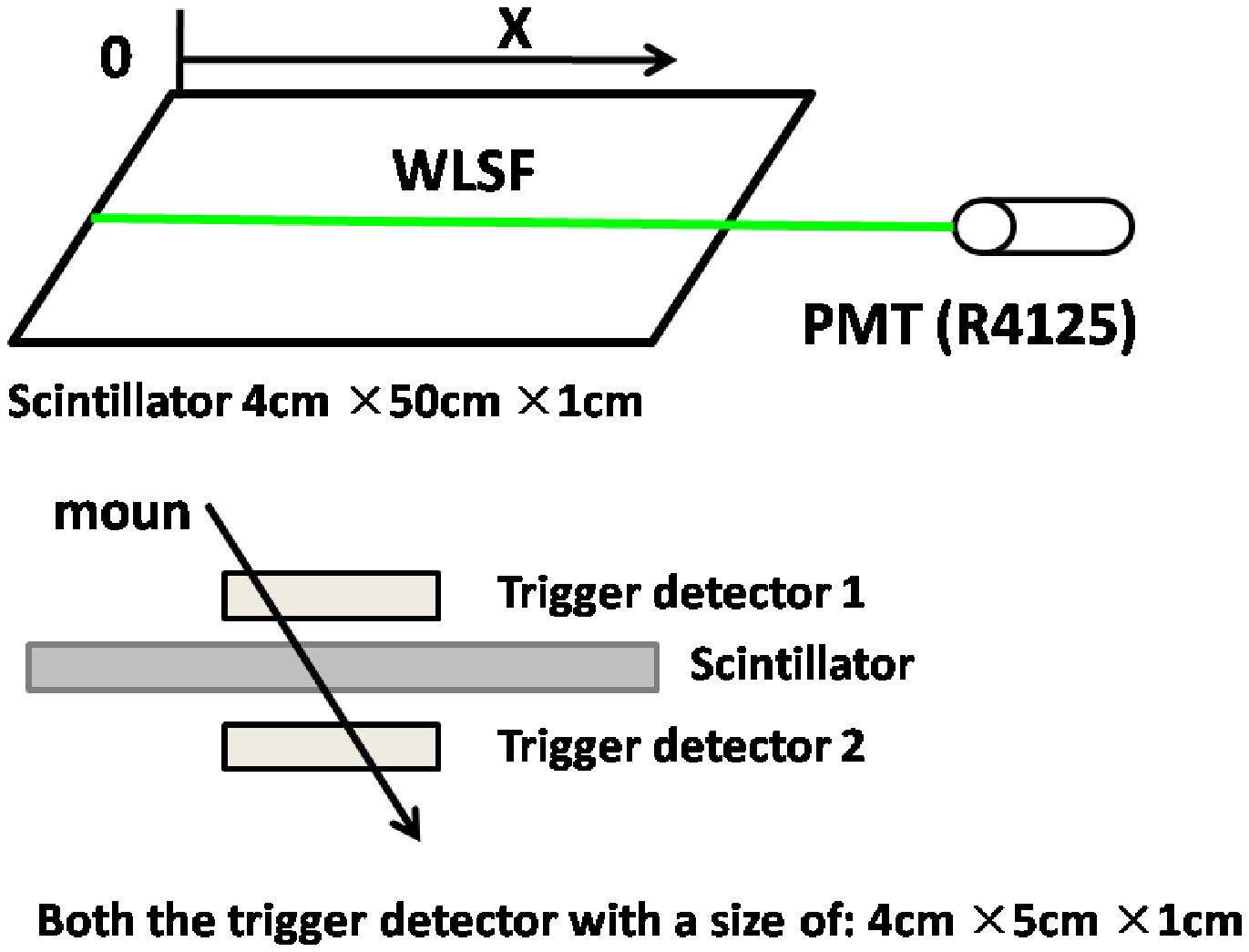}
\includegraphics[width=0.8\linewidth]{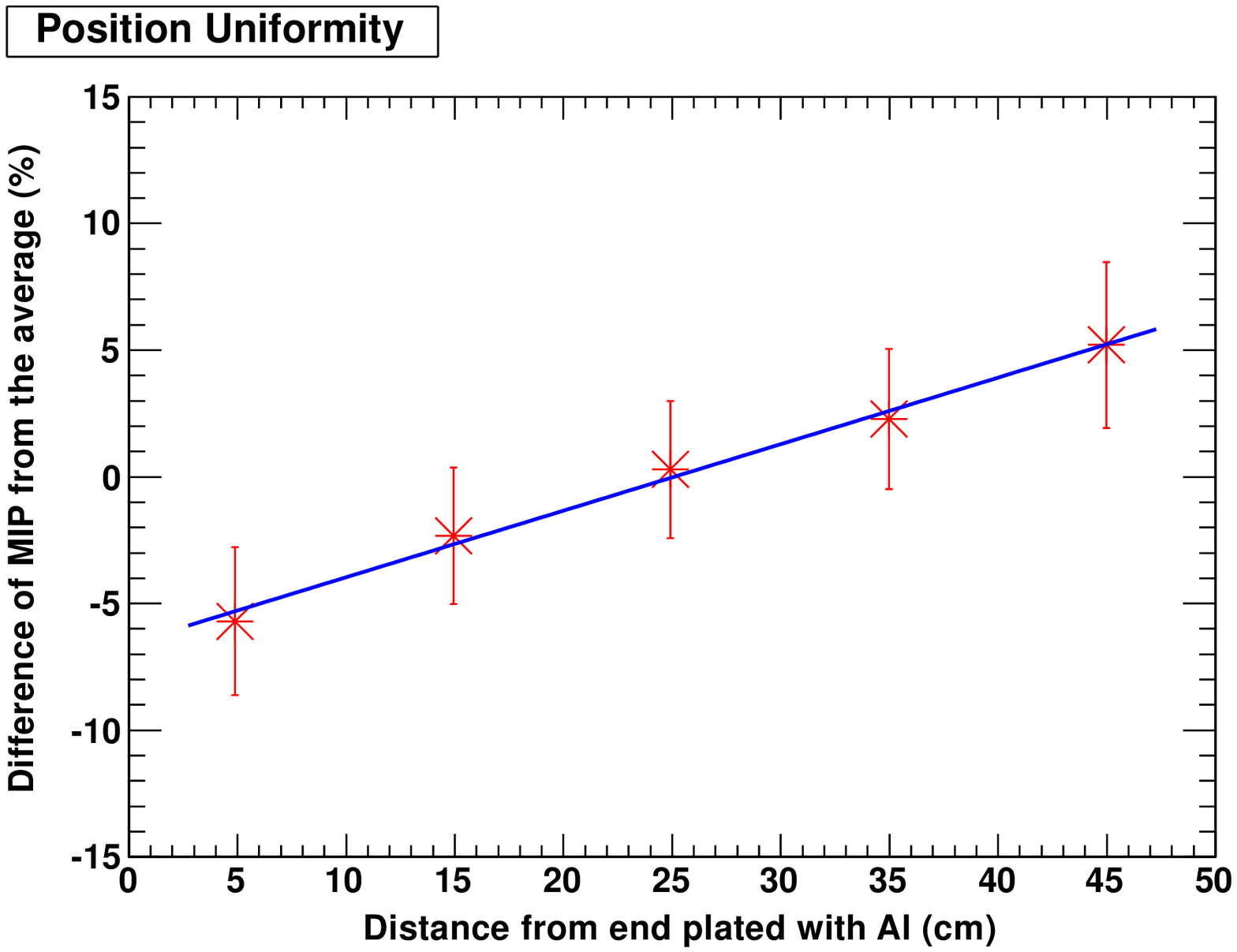}
\figcaption{Top: Experiment set-up for single muon measurement. Bottom: the uniformity of YAC-II detector.}
\label{figure-07}
\end{center}

\subsection{Linearity of PMT plus scintillator}
The linearity and the saturation of the plastic scintillator and PMT used in the YAC detector have also been
calibrated by using the accelerator beam of the BEPCII (Beijing Electron Positron Collider, IHEP, China).
The accelerator-beam experiment shows a good linearity between the incident particle flux and YAC-II output
below $5\times10^6$ MIPs and the saturation effect of the plastic scintillator satisfies YAC detector's requirement.
The details are described in ~\cite{Hardware-YAC-chen}.

\subsection{The correlation between high-gain PMT and low-gain PMT }
 Since we can not measure a single muon  using low-gain PMT (R5325) directly, we calibrate the low-gain PMT output
 using the the correlation between high-gain PMT (R4125) and low-gain PMT (R5325) in their overlapping region.
 After the detectors running consistently for about 10 days, we get the gains of low-gain PMT (R5325) and high-gain PMT (R4125) of the same burst events which are recoded by ADC, as shown in Fig.~\ref{figure-08}. From Fig.~\ref{figure-08},
 it is found that there is a fine linear relationship between two PMTs data. We can then obtain the ratio (just the slope of the line in Fig.~\ref{figure-08}) between them. With a further calculation it may be possible to get ${N_b}$ only from output signals from low-gain PMT.

\begin{center}
\includegraphics[width=0.8\linewidth]{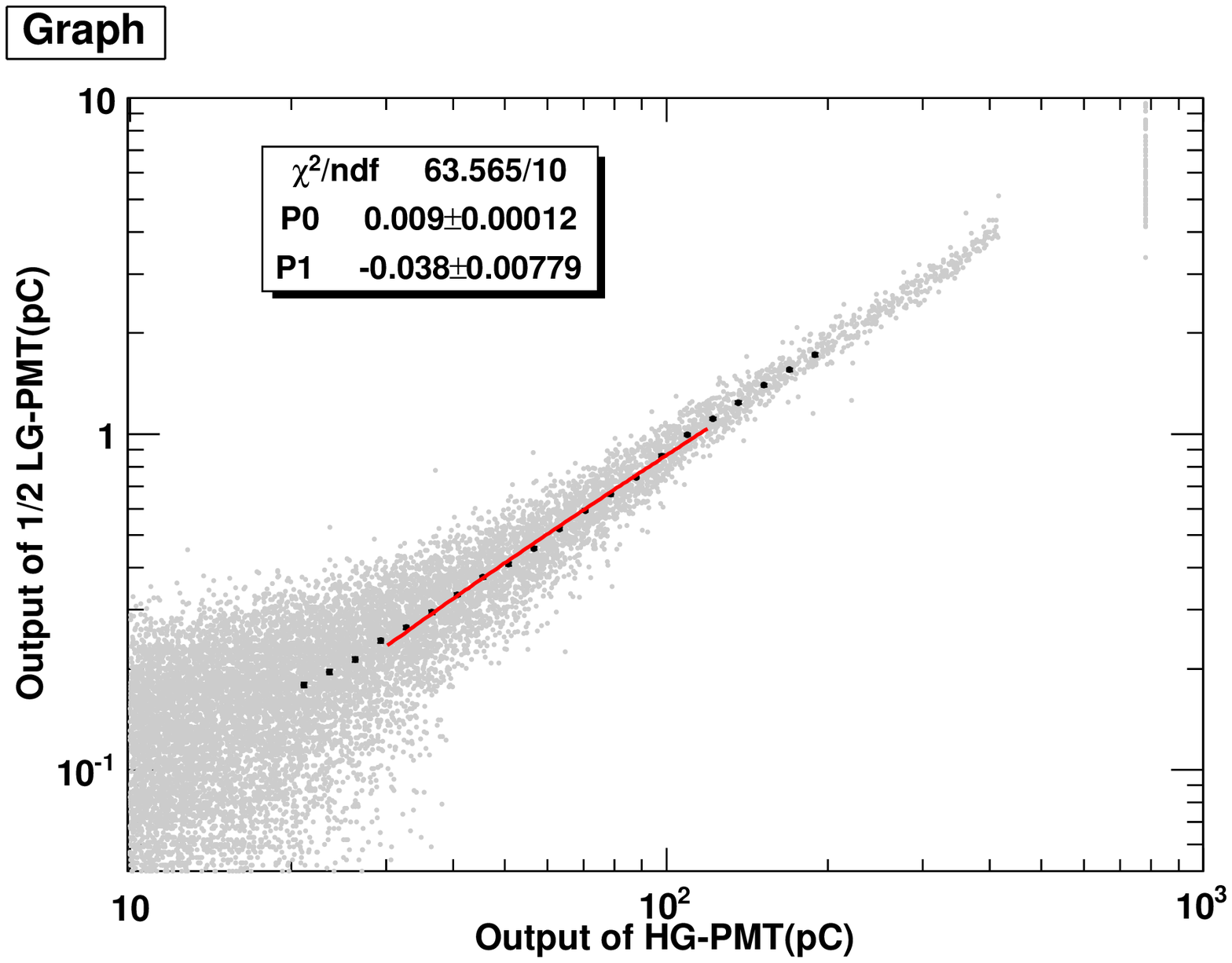}
\figcaption{The correlation between output signals from high-gain PMT (R4125) and low-gain PMT (R5325) obtained during 10 days operation of YAC-II.}
\label{figure-08}
\end{center}

\subsection{Consistency of the gains of different detector units}
Setting a lower detection threshold (say, taking 30 mV as the discrimination threshold) and ``any 1" (at least any one unit of the 124 units should be fired) as the trigger condition, YAC-II was operated for 10 days to measure the $N_b$-spectrum of each detector unit. Some differences are found from the $N_b$-spectrum of each unit. By slightly adjusting the working voltage of PMTs, consistent spectra are obtained. In Fig.~\ref{figure-09}, it is shown that the difference of the $N_b$-spectrum between 124 YAC detectors is less than 20\%.

\begin{center}
\includegraphics[width=0.8\linewidth]{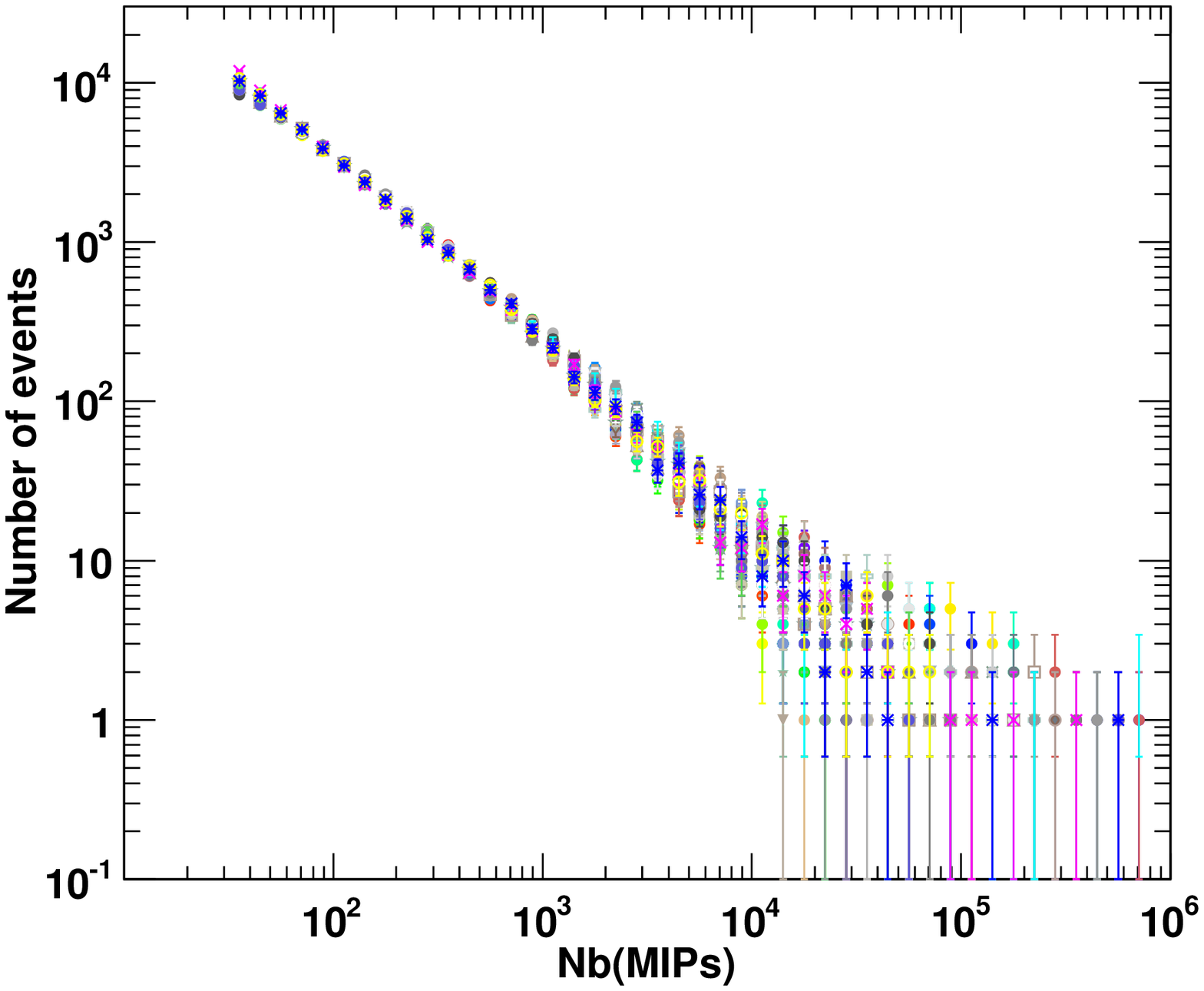}
\figcaption{The $N_{b}$-spectrum with the experimental data of YAC-II in 10 days. It is shown that the difference of
the $N_b$-spectrum between 124 YAC detectors is less than 20\%.}
\label{figure-09}
\end{center}

\section{Summary}
 A new air-shower core detector (YAC-II) has been set up at Yangbajing and started data taking from March, 2014. Each detector of YAC-II consists of 0.4 m$^2$ plastic scintillator, equipped with a high-gain PMT and a low-gain PMT which is put under 3.5 cm lead plate. The dynamic range of detector response is from 1 MIP to $10^6$ MIPs. Using the wave length shifting fiber to collect the scintillating light guarantees geometrical uniformity of the YAC detector response within 6\%. This new experimental condition improves the statistics of the high energy core event compared with Tibet-EC experiment by a factor of 100. Using the trigger conditions: at least anyone detector fired with detection threshold larger than 30 mV, the event rate reaches 3.5 Hz with dead-time rate 1\%. A study of Monte Carlo simulation shows that it is feasible to study the primary cosmic-ray composition using YAC-II array~\cite{YAC-liujs,YAC-lmzhai}.

\section{Acknowledgements}
The authors would like to express their thanks to the members of the Tibet AS$\gamma$ collaboration for the fruitful discussion. This work is supported by the Grants from the National Natural Science Foundation of China (Y11122005B, Y31136005C, Y0293900TF and 11165013) and the Chinese Academy of Sciences (H9291450S3 and Y4293211S5) and the Key Laboratory of Particle Astrophysics, Institute of High Energy Physics, CAS. The Knowledge Innovation Fund (H95451D0U2 and H8515530U1) of IHEP, China also provide support to this study.

%\footnote{Footnotes should be typeset in 8~pt  roman at the bottom of the page.}
\end{multicols}
\vspace{-1mm}
\centerline{\rule{80mm}{0.1pt}}
\vspace{2mm}

\begin{multicols}{2}

\end{multicols}

\clearpage

\end{CJK*}
\end{document}